# Evidence from web-based dietary search patterns to the role of B12 deficiency in chronic pain

Eitan Giat, Rheumatology Unit, The Autoimmune Center, Sheba Medical Center

Elad Yom-Tov, Microsoft Research Israel

## Abstract

Profound vitamin B12 deficiency is a known cause of disease, but the role of low or intermediate levels of B12 in the development of neuropathy and other neuropsychiatric symptoms as well as the relationship of eating meat and B12 levels is unclear. Here we use food-related internet search patterns from a sample of 8.5 million US-based people as a proxy to B12 intake and correlate these searches with internet searches related to possible effects of B12 deficiency.

Food-related search patterns are highly correlated with known consumption and food-related searches (Spearman 0.69). Awareness of B12 deficiency was associated with a higher consumption of B12-rich foods and with queries for B12 supplements.

Searches for terms related to neurological disorders were correlated with searches for B12-poor foods, in contrast with control terms. Popular medicines, those having fewer indications, and those which are predominantly used to treat pain are more strongly correlated with the ability to predict neuropathic pain queries using the B12 contents of food.

Our findings provide evidence for the utility of using Internet search patterns to investigate health questions in large populations and suggest that low B12 intake may be associated with a broader spectrum of neurological disorders than currently appreciated.

Vitamin B12 has long been known to be a cause of megaloblastic anemia and neurological disorders [1]. Very low levels of B12 cause subacute combined degeneration of the dorsal and lateral spinal columns, manifested by paresthesia, ataxia, impaired sense of vibration and proprioception, progressive weakness, spasticity and paraplegia. These symptoms may be irreversible, depending on duration. Other neurological abnormalities may be caused by B12 deficiency, such as peripheral neuropathy, memory loss, irritability and dementia [2,3], but the association of these symptoms with serum B12 levels and B12 intake is unclear. Many cases of B12 deficiency come to medical attention because of these symptoms, but seemingly asymptomatic B12 deficiency is very common, with prevalence up to 10-25% [4]. Furthermore, neurological symptoms due to B12 deficiency may appear in patients within a normal range of serum B12 levels [5].

There is no clear cut-off for serum B12 levels [6]. Levels higher than 300 pg/ml are considered normal, between 200 and 300 pg/ml are considered borderline, and less than 200 pg/ml are considered low. Measurements of serum B12 levels are not always reliable, for a variety of reasons [6]. Several methods are used by different labs, resulting in different normal ranges. Furthermore, results are highly variable, with absolute intra-individual variation >100 pg/ml on repeat testing in one fifth of the patients. Serum B12 concentration may be normal in 5% of patients with documented B12 deficiency [7].

Intermediate or low levels of B12 may contribute to the development of symptoms that are not typically associated with B12 deficiency. For instance, B12 dietary intake inversely correlated with sleep duration in 1902 healthy Japanese patients [8]. B12 levels inversely correlated with cardiovascular autonomic neuropathy in diabetics (mean level of 289 pg/ml), suggesting that intermediate levels may also be clinically significant.  Some have suggested low B12 levels may be a risk factor for depression, but evidence is conflicting [9,10].

The effect of diet on B12 levels is not fully understood and depends on food source and bioavailability. Usual western diet contains between 5-7 mcg of daily B12 intake, which is similar to the daily recommended intake, while body stores of B12 are 2000-5000 mcg [6]. Foods derived from animals are the major dietary sources of B12, but supplements containing B12 are commonly used in the United States, and many foods are fortified with B12 (such as non-dairy milks, meat substitutes, and breakfast cereals) [4]B12 is naturally protein-bound and its absorption depends on gastric acidity and intrinsic factors, which may be influenced by drugs and comorbidities, but reduced absorption of unbound B12 found in vitamin supplements is also common among healthy subjects.  Evidence on the effect of B12 intake from meat is confounding. A study assessing 2999 subjects from the Framingham Offspring study found meat intake did not affect B12 levels, and suggested a protective effect from supplements and B12 fortified foods [4].  In Norway, where cereals are not fortified with B12, dietary intake of B12 from milk and fish was associated with higher plasma vitamin B12 concentrations, whereas B12 from meat or eggs intake was not [11]. In contrary, analysis from the Dutch B-PROOF study [12] suggests that the bioavailability of B12 intake from meat is comparable to B12 intake from milk. To our knowledge, B12 intakes from different types of meat, such as beef, pork or chicken have not been compared.

As mentioned above, many symptoms typically associated with B12 deficiency can occur among patients with intermediate to normal levels of B12. We therefore sought to assess the impact of diet on these symptoms. Internet searches have been shown to reflect behavior in the physical world [13], as people use Internet search engines for information regarding many everyday behaviors, thus reflecting their diet, habits, and illnesses [14]. Here we analyzed internet searches related to food to quantify dietary B12 intake from different sources of food and their association with B12-related symptoms. We also sought to find a connection between B12 intake and other symptoms that are not typically associated with B12 deficiency, such as chronic pain or fibromyalgia.

## Methods

### Data collection

A list of 212 target terms which are possibly related to different types of chronic pain, in 8 categories: Antidepressants, neuropathic drugs, other pharmaceuticals (antihistamines and anti-acids), descriptions of pain, disorders related to excess acid, over the counter antacids, psychotherapy, and medical cannabis. Generic and commercial drug names of these disorders were taken from three websites: Uptodate, Micromedex and drugs.com. We also tested 27 medical terms (pharmaceuticals and conditions) which were unlikely to be associated with B12 deficiency and use these as control terms. Control terms were chosen from an online medical dictionary (Medterms.com). To ensure randomness, we chose the first terms in the alphabetical list, omitting terms which were rarely searched and could therefore can bias the results.

We extracted all searches made in English by people in the USA on the Bing search engine. For each search, we extracted an anonymized user identifier, the text of the search and the zip code from which the user made the search. Searches were categorized into three classes:

1. Recipe searches: Searches that contained a reference to a food recipe, as identified by a propriety classifier.

2. Term searches: Searches that contained one or more of the target terms (212 terms above), or the 27 control terms.

3. B12 searches: Searches that contained the term B12 in them.

To identify ingredients required for each recipe we used the list of recipes from Recipes Wikia

(http://recipes.wikia.com/wiki/Recipes_Wiki, available at:

http://recipes.wikia.com/wiki/Special:Statistics). Specifically, we extracted all recipes that contained one

or more of the following 12 food ingredients, which have a variety of B12 contents: Shellfish, Mackerel,

Trout, Salmon, Tuna, Pig, Beef, Turkey, Chicken, Egg, Milk, and Tomatoes. A total of 9449 recipes

contained one or more of these ingredients.

Additional data required for analysis included the following:

1. B12 contents per serving of the 12 food ingredients, from the US National Institues of Health
   Office of Dietary Supplements (https://ods.od.nih.gov/factsheets/VitaminB12-
   HealthProfessional/). See **Table** 1.

2. Food spending data, from the US Bureau of Labor Statistic's Consumer Expenditure Survey 2015
   (https://www.bls.gov/cex/tables.htm)

3. Cost of food items, from the US Department of Agriculture's Center for Nutrition Policy and
   Promotion (https://www.cnpp.usda.gov/data).

4. Food consumption data, from the US Department of Agriculture's Economic Research Service
   (https://www.ers.usda.gov/data-products/food-availability-per-capita-data-system/)

5. The number of indications for each medical drug, from the Canadian Vigilance Adverse
   Reactions Online database (http://www.hc-sc.gc.ca/dhp-mps/medeff/databasdon/). All
   indications reported by at least 10 cases were considered.

The study was deemed exempt by the Microsoft Institutional Review Board.

## Results

Approximately 8.5 million people searched for a target medical term, food recipe, or B12 were included in the analysis. Of 212 target medical terms, only 101 terms that were queried by 1000 or more people were included for analysis. A total of 3297 recipes were matched to at least one query. Manual examination of a random sample of 400 queries found that 87.8% of queries were correctly matched to a recipe.

We use searches for food recipes as a proxy for consumption, as suggested in West et al. [15]. To further validate that this proxy is accurate, we correlated the fraction of known expenditure on 6 food items (beef, pork, poultry, fish and seafood, eggs, and milk) at each of 4 US regions, with the number of queries for recipes containing each of these food items, weighted by their cost.  The correlation between the two is 0.69 (P=0.0002, N=24), suggesting that searches for foods are strongly correlated with consumption. Thus, we posit that B12 consumption can be estimated through recipe searches.

Awareness of B12 deficiency is also correlated with consumption. The estimated B12 consumption of users was computed by summing the multiplication of the B12 contents per serving of each of the 12 food types by the number of times a user searched for about each food type. The average value of the estimated consumption for people who asked about B12 was 2.407mcg, compared to 2.395mcg for those who did not (0.5% difference, statistically significant: ranksum test, $P<10^{-10}$). However, the difference in estimated consumption for the 4701 people who specifically asked about B12 deficiency was 2.219mcg, compared to 2.395mcg for those who did not (8% difference, ranksum $P<10^{-10}$). We interpret this finding as additional supporting evidence for the claim that estimating B12 consumption through searches for recipes is valid, given the lower estimated B12 consumption for people who indicated an awareness of this deficiency.

The vast majority (91.8%) of people in our sample did not ask about B12 deficiency, supplements, or serum level. However, among people who asked about B12 deficiency, 3.8% (180 of 4701) asked about B12 supplements (either in the form of dietary supplements or injections), compared to 0.03% in the population who didn't ask about B12 deficiency. Thus, approximately 151 more times people who asked about deficiency asked about supplements.

To estimate the correlation between food consumption and the likelihood of asking about each of the medical terms, we constructed an individual-level model where each person is represented through the number of searches they made for recipes containing each of 12 food items and, separately, as to whether they asked about each of the target terms. The two behaviors were linked through a linear classification model where the independent terms were the number of searches for each food item, and the dependent variable were whether a person asked about the medical term.

**Table** 2 shows the 10 medical terms for which the individual-level model reached the highest goodness of fit ($R^2$) value (denoted by $R^2$I), together with the correlation between model coefficients (per food item) and the B12 contents per serving of each food item, denoted by CoB12. $R^2$I values are low even for the best fitting models, but their median value is 78 times greater than $R^2$I values for the control terms (medical terms unassociated with B12 deficiency) (for the top 10 terms in each list, ranksum, P=0.002). We further note that $R^2$I is correlated with the number of people who asked about each medical condition (Spearman 0.86, P<$10^{-5}$). Thus, conditions we hypothesized may be linked to B12 deficiency can indeed be associated with food consumption, especially for the more common medical conditions.

Next, we used the models constructed above to estimate the contribution of B12 contents to each medical condition. The correlation between model coefficients and B12 contents of food items (CoB12) is negative for all terms in Table 2, indicating that people searching for B12-rich foods are less likely to search for the medical terms, compared to people searching for B12-poor foods.

Different CoB12 values observed across medical terms are explained by several factors. First, the average value of the regression coefficients for all but tomatoes were at least 2.3 times greater than those of tomatoes, whose B12 contents is negligible. Second, the Spearman correlation between $R^2I$ and CoB12 is -0.29 (P=0.004, N=101), whereas for the control terms it is non-significant (Spearman 0.35, P=0.08, N=27). Thus, among the target terms, coefficients of better fitting models are larger (more negative) than of poorly fitting models. Finally, CoB12 for medical drug terms is correlated with drug indications, as described below. On average, the correlation for the control terms is only 0.04, whereas the same number for the target medical terms was, on average -0.31 (ranksum, 0.01). Moreover, positive correlations were only observed for individual control terms, and never among the target terms.

The number of possible indications of a drug can affect its correlation with CoB12, since indications to unrelated conditions may mask the true correlation. We identified the number of indications from the Canadian Adverse Reactions database. Indications were further stratified to identify those indications that contained the term "pain". The number of indications is strongly correlated with the number of people who asked about a medical condition (Spearman 0.53, P=$10^{-4}$, N=95), but the number of pain indications only marginally significantly correlated with the number of people (Spearman 0.29, P=0.048, N=95). These correlations are likely due to broad-spectrum drugs being offered to treat more conditions than drugs which can only be prescribed to specific indications.

Next, we tested the correlation between the broadness of indication of a drug on CoB12. We modeled CoB12 of medical drugs as a function of the interaction between the number of people who asked about a term, the number of indications and the number of pain-related indications using a rank regression model, weighting samples by $R^2I$. The model reaches an $R^2$ of 0.48 (P=$10^{-6}$, N=49). Statistically significantly correlated coefficients of the model are:

1. The number of indications: Slope is -0.43 (P=$10^{-6}$), meaning that fewer indications are correlated with higher CoB12.

2. The number of pain indications: Slope is 0.20 (P=0.04), meaning that the more pain-related indications, the higher CoB12.

We interpret the first as showing that CoB12 of non-specific medicines is lower, as the effect of B12 deficiency is masked by people with other medical conditions. The second correlation is interpreted as showing that the more a drug is used to treat pain (ostensibly caused partly by B12 deficiency), the better the correlation to B12 consumption.

Taken together, our results indicate that the more focused a term is in its medical application and the more popular, the better the model is in estimating the correlation of B12 consumption with its appearance. The more people consume B12-rich foods, the less likely they are to query for those medical terms we hypothesized are related to B12 deficiency.

## Discussion

Our study is the first to employ analysis of internet searches to estimate B12 intake and its effect. We here show a strong correlation between food consumption and internet searches, indicating that searching for recipes reflects food consumption. Our analysis allows an assessment of B12 intake from different types of foods, and especially meats. The Framingham Offspring study [4] found that B12 intake from meat did not change B12 levels, but this study regarded all types of meat as one entity, whereas our data strongly suggests the different types of meat provide different B12 intakes, and therefore should be analyzed separately. Most physicians are aware of the risk of B12 deficiency in strict vegans, but the intake from an average non-vegetarian diet is considered sufficient in B12, with no preference to any type of meat. This is likely because an average non-vegetarian diet includes the recommended daily intake. Additionally, vitamin B12 stores are high, and are easy to replenish. Other causes limiting B12 absorption, such as gastric acidity and intrinsic factor deficiency, are considered much more important factors in B12 deficiency. As a result, physicians do not consider non-vegetarians to be at risk for B12 deficiency, with no discrimination between beef or poultry based diets. Our study is the first to assess the differences in B12 intake from different sources of meat in daily diet, and suggests that the distinction between vegetarians and non-vegetarians may be inadequate for evaluating the risk for B12 deficiency.

Our data shown that some people are aware of B12 deficiency, and that this is reflected in their diet. However, not all patients with B12 deficiency are aware of this shortage. Indeed, difficulties in assessing B12 levels may cause both patient and clinician to be unaware of its shortage. We here show that low B12 intake is correlated with symptoms associated with B12 deficiency among people who did not search for B12-rich foods, supposedly unaware of B12 deficiency. We performed most of our analyses on a population of people who did not express awareness of B12 deficiency or sufficient intake thereof because patients who are aware of B12 deficiency tend to use supplemental B12, and this may mask the effect of dietary B12 intake. Besides further emphasizing the importance of dietary trends, these results suggests that a significant population of patients may suffer from symptomatic B12 deficiency without being aware of the cause of their symptoms. To date, studies evaluating the effect of B12 intake only measured B12 serum levels as an outcome. Our results suggest that other outcomes, such as peripheral neuropathy and other neurological complaints should also be considered.

Our study found a correlation between B12 intake and different drugs and diseases. Paresthesia, which is neuropathic in origin, may be a symptom of B12 deficiency. Not surprisingly, the four strongest correlations are related to neuropathy and pain and its treatment. Gabapentin and Lyrica are common treatments for neuropathic pain or central neuropathic pain. Tramadol is also prescribed for neuropathic pain as well as other types of pain. Internet searches for sertraline, citalopram, duloxetine and trazodone were also inversely correlated with the estimated B12 intake. These drugs are common anti-depressants, and our results strengthen the yet controversial association between low B12 levels and depression [16]. Searches for oxycodone, a potent opiate painkiller, were also associated with lower B12 dietary intake. Oxycodone is not the drug of choice for neuropathic pain, but is usually reserved for severe or refractory pain. B12 is not typically associated with non-neuropathic pain, though parenteral B12 has been shown to alleviate low back pain [17]. The association of B12 intake and oxycodone suggests that B12 may have a role in severe or refractory pain which is not neuropathic. Interestingly, omeprazole, an important proton pump inhibitor was also correlated with a decreased B12 dietary intake. The reason for this relationship is unclear.  Omeprazole, through its gastric acid lowering effects, is considered to be a cause for B12 deficiency, but B12 deficiency is not known to cause gastric acidity. A possible explanation might be that peptic pain may be amplified in B12 deficiency, but this has never been shown or looked into.

Our study has several limitations. First, a search for recipe does not necessarily translate to the precise personal consumption and may be biased. We validated our results at the population level (regional consumption), but individual consumption may differ. Past work has found a high correlation between recipe searches and actual consumption [15]. Moreover, in our study we require that recipe searches correlate with the profile of food consumption (i.e., the fraction of beef over chicken), not the actual rates of individual consumption. Additionally, though correlations were not very high, the fact that despite this limitation we were able to show differences in the use of the target terms as a function of B12 content (but not in searches unrelated to B12), strengthens the notion that differences in B12 intake from different types of meat are important in determining the risk for B12 deficiency and related symptoms.

The use of internet searches as a proxy for health conditions may also be biased. For example, B12 has been suggested to be associated with depression, but the depression may cause a decrease in internet use, making it difficult to identify this disorder through internet queries.

Another limitation is our inability to quantify the impact of dietary B12 on serum levels. Our model identified an increase of risk with consumption of meat with low B12 content, but it cannot predict the serum B12 levels for a specific diet. On the other hand, the difficulties in relying on B12 levels and the occurrence of clinical B12 deficiency alongside seemingly normal serum B12 levels, suggest that assessing symptoms may be more important than estimating serum levels.

Despite these limitations, utilizing internet searches as a proxy for B12 intake has the advantage of overcoming memory bias and has the ability to go into dietary details, which is much more difficult to perform in questionnaire based studies.

Altogether, our data suggests that meat alone is not sufficient to prevent B12 deficiency and that the source of meat should also be considered. Our data also suggests that B12 intake inversely correlates with neurological symptoms, implying a role for B12 among a seemingly unaware population. Physicians should be aware of the possible role of B12 in any patient with neurological complaints or unexplained pain. Sure enough, further research is necessary to confirm and determine the clinical significance of our results.

Table 1. B12 content in different types of meat. This table is based on

https://ods.od.nih.gov/factsheets/VitaminB12-HealthProfessional/

| Meat source | B12 content (mcg/100 grams) |
|---|---|
| shellfish | 98.89 |
| Mackarel | 19 |
| Beef | 6 |
| Trout | 3.8 |
| Salmon | 2.4 |
| Tuna | 1.6 |
| Milk | 0.9 |
| Turkey | 0.8 |
| Egg | 0.6 |
| Pork | 0.4 |
| Chicken | 0.3 |

Table 2: $R^2$ of the individual-level models to predict the likelihood of asking about medical terms given questions about foods (denoted by $R^2I$) and the correlation between model coefficients and B12 contents of these food items, for the 10 medical terms with the highest $R^2I$ values.

| Medical term | $R^2I$ | CoB12 |
|---|---|---|
| gabapentin | 0.003 | -0.405 |
| tramadol | 0.003 | -0.473 |
| neuropathy | 0.002 | -0.322 |
| lyrica | 0.002 | -0.398 |
| omeprazole | 0.002 | -0.397 |
| sertraline | 0.002 | -0.303 |
| citalopram | 0.002 | -0.333 |
| oxycodone | 0.002 | -0.269 |
| duloxetine | 0.002 | -0.329 |
| trazodone | 0.002 | -0.354 |